\documentstyle[11pt,epsf]{article}
\begin{document}
\newcommand{\largepspicture}[1]{\centerline{\setlength\epsfxsize{13cm}\epsfbox{#1}}}
\newcommand{\vlargepspicture}[1]{\centerline{\setlength\epsfxsize{15cm}\epsfbox{#1}}}
\newcommand{\smallpspicture}[1]{\centerline{\setlength\epsfxsize{7.5cm}\epsfbox{#1}}}
\newcommand{\tinypspicture}[1]{\centerline{\setlength\epsfxsize{7.4cm}\epsfbox{#1}}}
\newcommand{\ewhxy}[3]{\setlength{\epsfxsize}{#2}
            \setlength{\epsfysize}{#3}\epsfbox[0 20 660 580]{#1}}
\newcommand{\ewxy}[2]{\setlength{\epsfxsize}{#2}\epsfbox[10 30 640  590]{#1}}
\newcommand{\ewxynarrow}[2]{\setlength{\epsfxsize}{#2}\epsfbox[10 30 560 590]{#1}}
\newcommand{\ewxyvnarrow}[2]{\setlength{\epsfxsize}{#2}\epsfbox[10 30 520 590]{#1}}
\newcommand{\ewxywide}[2]{\setlength{\epsfxsize}{#2}\epsfbox[0 20 380 590]{#1}}
\newcommand{\err}[2]{\raisebox{0.08em}{\scriptsize{$\hspace{-0.2em}\begin{array}{@{}l@{}}
                     \plus\makebox[0.55em][r]{#1}\\[-0.05em]
                     \minus\makebox[0.55em][r]{#2}
                     \end{array}$}}}
\newcommand{\plus}{\makebox[15pt][c]{$+$}}
\newcommand{\minus}{\makebox[15pt][c]{$-$}}
\newcommand{\mpr}{\frac{m_{\pi}}{m_{\rho}}}

\newcommand{\eq}{eq.~}
\newcommand{\beq}{\begin{equation}}
\newcommand{\eeq}{\end{equation}}
\newcommand{\bea}{\begin{eqnarray}}
\newcommand{\eea}{\end{eqnarray}}
\newcommand{\nn}{\nonumber}

\newcommand{\oks}{{1 \over \kappa_{\rm sea}}}
\newcommand{\okl}{{1 \over \kappa_{\rm sea}^{\rm light}}}
\newcommand{\okss}{{1 \over \kappa_{\rm sea}^{\rm strange}}}
\newcommand{\okst}{{1 \over \kappa^{\rm strange}}}
\newcommand{\okv}{{1 \over \kappa_{\rm V}}}
\newcommand{\okvc}{{1 \over \kappa_{\rm V}^c}}
\newcommand{\okvl}{{1 \over \kappa_{\rm V}^{\rm light}}}
\newcommand{\okc}{{1 \over \kappa_{\rm sea}^c}}

\newcommand{\kss}{\kappa_{\rm sea}^{\rm strange}}
\newcommand{\kst}{\kappa^{\rm strange}}
\newcommand{\ks}{\kappa_{\rm sea}}
\newcommand{\kv}{\kappa_{\rm V}}
\newcommand{\kvc}{\kappa_{\rm V}^c}
\newcommand{\kvl}{\kappa_{\rm V}^{\rm light}}
\newcommand{\ksc}{\kappa_{\rm sea}^c}
\newcommand{\kc}{\kappa_{\rm sea}^c}
\newcommand{\ksl}{\kappa_{\rm sea}^{\rm light}}
\newcommand{\kl}{\kappa_{\rm sea}^{\rm light}}

\newcommand{\mq}{m}
\newcommand{\mqv}{m^{\rm V}}
\newcommand{\mqs}{m^{\rm sea}}
\newcommand{\mql}{m^{\rm light}}
\newcommand{\mqst}{m^{\rm strange}}

\newcommand{\mot}{m_{\sf ss}}
\newcommand{\mtf}{m_{\sf vv}}
\newcommand{\moth}{m_{\sf sv}}

\newcommand{\mps}{m_{\rm PS}^2}
\newcommand{\mpsot}{m_{\rm PS, {\sf ss}}^2}
\newcommand{\mpstf}{m_{\rm PS, {\sf vv}}^2}
\newcommand{\mpsoth}{m_{\rm PS, {\sf sv}}^2}

\newcommand{\mv}{m_{\rm V}}
\newcommand{\mvot}{m_{\rm V, {\sf ss}}}
\newcommand{\mvtf}{m_{\rm V, {\sf vv}}}
\newcommand{\mvoth}{m_{\rm V, {\sf sv}}}

\newcommand{\chisq}{\chi^2/{\sf d.o.f}}

\title{\vskip -3cm \makebox[9.5cm]{}{\normalsize WUB 97-14} \\
\vspace{-0.2cm}\makebox[9.5cm]{}{\normalsize HLRZ 1997\_16} \\ 
\vspace{1.6cm}
{\huge{\sf{Light Quark Masses with Dynamical Wilson Fermions}}}}
\vspace{0.2cm}
\author{\sf{SESAM-Collaboration:} \\ 
\normalsize{
N.~Eicker$^{\rm a}$, U.~Gl\"assner$^{\rm b}$, S.~G\"usken$^{\rm b}$,
H.~Hoeber$^{\rm b}$
}
\\ 
\normalsize{
P.~Lacock$^{\rm a}$, Th.~Lippert$^{\rm a}$ G.~Ritzenh\"ofer$^{\rm a}$, K.~Schilling$^{\rm
  a,b}$,
}
\\
\normalsize{
G.~Siegert$^{\rm a}$, A.~Spitz$^{\rm a}$, P.~Ueberholz$^{\rm b}$ and J.~Viehoff$^{\rm b}$ 
}
\vspace{0.2cm}\\
\small{
{\rm $^a$}HLRZ c/o Research Center J\"ulich, D-52425 J\"ulich and DESY, D-22603 Hamburg, Germany,}\\
\small{
{\rm $^b$}Physics Department, University of Wuppertal, D-42097 Wuppertal, Germany.}
}
\date{}
\maketitle
\begin{abstract}
We determine the masses of the light and the strange quarks in the
$\overline{MS}$-scheme using our high-statistics lattice simulation of QCD with
dynamical Wilson fermions. For the light quark mass we find $m^{\rm light}_{\overline{MS}}(2\, {\sf GeV}) =
2.7(2)\, {\sf MeV}$, which is lower than in quenched simulations.
 For the strange quark, in a sea of two dynamical
light quarks, we obtain $m^{\rm strange}_{\overline{MS}}(2\, {\sf GeV})
= 140(20)\, {\sf MeV}$.
\end{abstract}
%
\section{Introduction}
The masses of $u$, $d$, and $s$ quarks constitute fundamental
parameters of the Standard Model.  Phenomenologically, however, they 
remain among the most poorly known quantities within  its
scenario.

While lowest order chiral perturbation theory offers a fairly easy
access to the determination of {\it ratios of quark masses} from
the empirical mesonic spectrum\cite{leutwyler}, one has to apply 
much more meticulous techniques, such as QCD sum rules\cite{sumrules}
or lattice QCD\cite{mackenzie,gupta}, in order to arrive at {\it absolute
  values}. At this stage, however, these two methods appear to lead to
contradictory results. 

In practice, both of these approaches carry their specific merits
and shortcomings. While the sum rule results are sensitive to the
choice of parametrizations, as elaborated in ref.\cite{guptaneu}, the
lattice results have been established so far for pure gluon dynamics
only\cite{mackenzie,gupta}. 

It is therefore of considerable interest to study the dynamical
effects of vacuum fluctuations, originating from light quarks, onto
light quark properties such as their masses.  This holds in particular
for Wilson-like discretizations of fermions which, unlike staggered
fermions, are free of flavour symmetry violations on the
lattice. Until recently, however, computing resources and techniques were too limited
to allow for the generation of reliable samples of vacuum
configurations with appropriate statistics. Nevertheless, a recent
rough analysis of world data\footnote{ including a preliminary data
  set from SESAM.} seems to suggest\cite{gupta} that unquenching from
$N_f =0$ to $N_f = 2$ might have a sizeable impact on the value of the light
and strange quark masses, lowering them by as much as 50 \%.

In this letter, we present a lattice analysis for the light and
strange quark masses based on our measurements of
the pseudoscalar and vector mesons determined in a sea of two degenerate
dynamical quarks. We have generated three sets of gauge configurations
at three sea-quark masses but at the same coupling; 
each set comprises 200 independent gauge configurations. 
Good signals in the autocorrelation functions and the
use of a blocking method give us confidence as to the reliability of
the quoted errors.

The masses of the light and the strange quark are extracted from the meson
data by fixing two mass ratios, ${M_\pi \over M\rho}$ for the light
and one of ${M_{\phi} \over M\rho}$, ${M_{K} \over M\rho}$ or ${M_{K^*} \over
  M\rho}$ for the strange; the lattice spacing is obtained from $M_\rho$. 

We identify the dynamical quarks with the degenerate doublet of 
isopsin symmetric quarks, $u$ and $d$, called light quark in the
following. Thus, using our data for the pseudoscalar and the vector at the three
sea-quarks, we can extract the mass of the light quark in a sea
of light quarks.

To simulate the strange quark we need to introduce valence quarks
that are unequal to the dynamical quarks: at each sea-quark we
evaluate meson masses with strange valence quarks and
perform an extrapolation of these masses to the physical sea of light
quarks. This procedure allows us to calculate the masses of the $K$, $K^*$ and the
$\phi$ with a dynamical light quark(for $K$ and $K^*$) and strange
quarks in a sea of light quarks. 

The definition of a strange quark mass in a sea of
light quarks requires an analysis of lattice data in terms of
sea and valence quark masses, for which we describe a suitable 
parametrization in section \ref{technical}. 
As an aside, we also comment on a possible flaw
in the extraction of quark masses from quenched data with Wilson-like fermions.

%
%
\section{Simulation details}

%
\begin{table}[tb]
\begin{center}
\begin{tabular}{|cccc|}
\hline
\hline
\multicolumn{4}{|c|} {$\beta_{\rm dyn} = 5.6$, $N_f = 2$, $16^3 \times
  32$}\\
\hline
$\kappa_{\rm sea}$ & 0.156 & 0.1570 & 0.1575 \\ 
\hline
number of configurations & 200 & 200 & 200 \\   
\hline
$\kv - \kv$ combinations & 15 & 15 & 15 \\
\hline
\hline
\multicolumn{4}{|c|} {$\beta_{\rm quenched} = 6.0$, $16^3 \times 32$}\\
\hline
\multicolumn{4}{|c|}{number of configurations: 200}\\
\hline
\multicolumn{4}{|c|}{$\kv - \kv$ combinations: 15}\\
\hline
\hline
\end{tabular}
\caption{Simulation details.
\label{simulation}}
\end{center}
\end{table}
%
In table \ref{simulation} we show the parameters of our simulation. In
addition to the dynamical simulation we performed a quenched study at
the matching quenched coupling\footnote{This is the quenched $\beta$
  which yields a similar lattice spacing.} to enable a direct comparison of full
and quenched QCD.
\par
At each of our three sea-quark values, characterized by the hopping
parameters $\ks$, we have investigated
zero-momentum two-point functions,
\beq
C_{AB}(t) = \sum_{\vec x}  \langle 0 | 
\chi_A^{\dagger}(0) \chi_B(x) | 0 \rangle \; ,
\end{equation}
for the pseudoscalar and vector particles, $\chi_{PS}(x) = \bar{q}'(x)
\gamma^5 q(x)$ and $\chi^{\mu}_{V}(x) = \bar{q}'(x)
\gamma^{\mu}q(x)$. We combined light-quark propagators with hopping
parameters equal and different to that of the corresponding sea-quark,
allowing for fifteen mass estimates per sea-quark. We use the
gauge-invariant Wuppertal-smearing procedure\cite{Wupsmear} to
calculate smeared-local and smeared-smeared correlators (with smearing
parameter $\alpha =4$ and with 50 iterations). Both types of
smearing are used to obtain mass-estimates by performing a
simultaneous single-exponential\footnote{We have checked that
  two-exponential fits yield stable ground state masses.} 
fit to the data on time-slices
10-15. Details will be given in \cite{SESspec}.
\par
In ref.\cite{SESauto}  we will present a detailed auto-correlation
analysis. We found integrated auto-correlation
times, $\tau_{\rm int}$, 
for the masses to lie around 25 HMC time-units with a slight
increase towards lighter sea-quarks. We have therefore chosen to
calculate propagators on configurations separated by 25 HMC
trajectories. To determine by how many units of
$\tau_{\rm int}$ the measurements need to be separated to achieve
complete decorrelation we performed a blocking analysis and
plotted the error as a function of the blocking size. 
At block size 6 (for $\ks = 0.156$ and $0.157$) and 7
(at $\ks = 0.1575$) we find the jackknife errors of the error to run into
plateaus\footnote{We consider this to be a conservative estimate.}. 
Consequently, these are the errors we will quote in the
following. A similar analysis for our quenched data shows no
increase in error with the blocksize (quenched configurations are
generated with an overrelaxed Cabbibo-Marinari heatbath update and are
separated by 250 sweeps).
\par
Errors (on the blocked data) are obtained using the bootstrap
  procedure. They correspond to 68 \% confidence limits of
  the distribution obtained from 250 bootstrap samples.
\section{Results - light and strange quarks\label{technical}}
The light quark mass is extracted by matching the ratio\footnote{We use the
  convention that masses\cite{PRD} in the continuum are denoted with
  capital ``$M$'', whereas lattice masses are written as ``$m$''.}: 
\beq
{m_{\rm PS} \over \mv} = {M_{\pi} \over M_{\rho}} = 0.1785 \, ,
\label{ratiolight}
\eeq
using data with (degenerate) valence quarks equal to the
sea-quarks. Generically, we call this data $\mot$. Owing to the fact
that we have but three different sea-quarks, 
\beq
\mqs = {1 \over 2}\left(\oks - \okc \right) \; ,
\eeq
we use linear extrapolations in the lattice quark mass:
\bea
\mpsot &=& c \left( \oks - \okc \right)\; , \label{pisym} \\
\mvot &=& m^{\rm crit} + b \mpsot \; .\nonumber 
\eea
%
\begin{figure}[tb]
\centerline{
\ewxynarrow{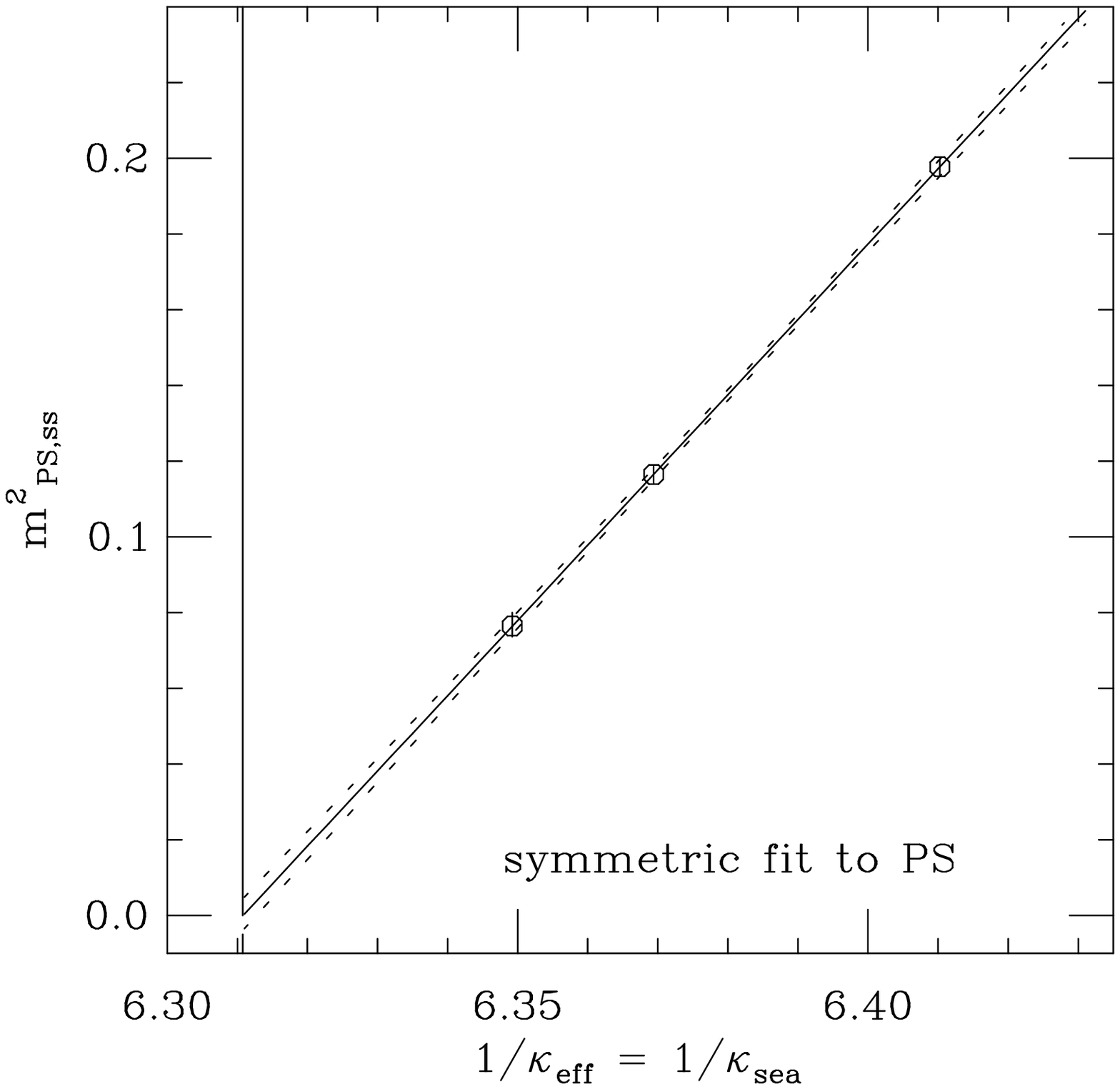}{250pt}       
\ewxynarrow{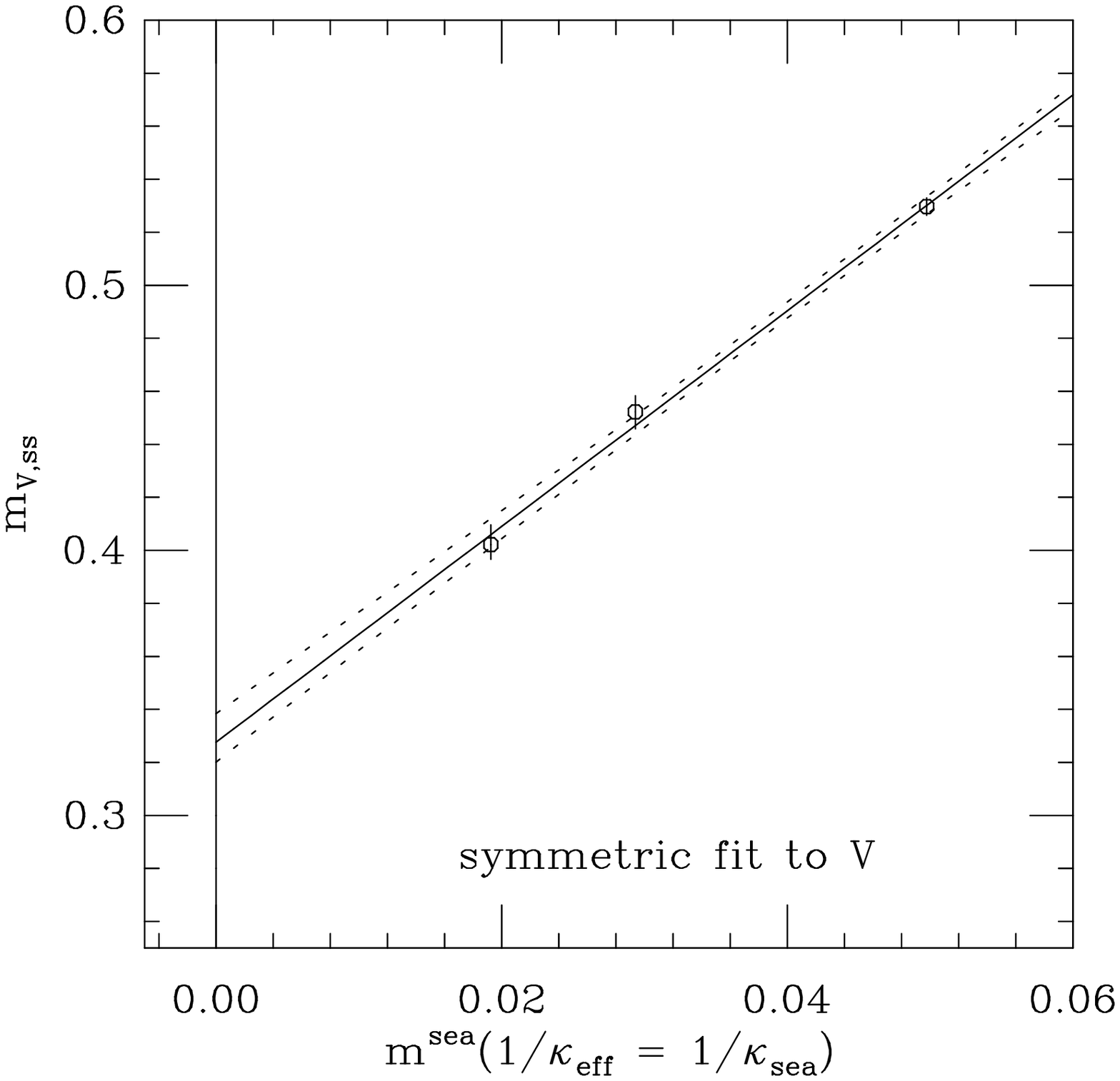}{250pt}
}
\caption{$\mpsot$ (left) as a function of $\oks$ and $\mvot$ as a
  function of $\mqs$ (in lattice units).
\label{fig1}}
\end{figure}
These fits, which we call ``symmetric'', are shown in figure
\ref{fig1}. We find the pseudoscalar mass to be extremely well matched by the
linear ansatz ($\chisq = 0.002$), whereas the vector masses may exhibit some
downward curvature ($\chisq = 1.1$). We find:
\begin{eqnarray}
&\ksc = 0.15846(5)\;\;\;\;  \ksl = 0.15841(5)\; , & \label{res1}  \\
& m^{\rm light} = {1 \over 2}\left(\okl - \okc \right) = 0.00088(6) \; .& \label{res2}
\end{eqnarray}
The corresponding lattice spacings from the rho are:
\bea
a^{-1}_{\rho} &=& 2.35(6)\;{\sf GeV} \;\;\;\;{\rm at }\;\; \ksc  , \\
a^{-1}_{\rho} &=& 2.33(6)\;{\sf GeV} \;\;\;\;{\rm at }\;\; \ksl .
\eea
We use the latter value to convert to physical units in the
$\overline{MS}$-scheme according to:
\beq
m_{\overline{MS}}(\mu) = Z_M(\mu a) \mqs a^{-1}_{\rho} \; ,
\eeq
with $Z_M(\mu a)$ calculated in boosted 1-loop perturbation theory
\cite{Zhang,Lepage} and run the values to 2 ${\sf GeV}$. Throughout,
we allow for a 3 \% uncertainty in $Z_M$. As a result we find:
\bea
m^{\rm light}_{\overline{MS}}(2\, {\sf GeV}) &=& 2.7(2)\, {\sf MeV} \; .
\eea
\vspace{0.5cm}\\
\par
As we have outlined in the introduction, the treatment of the
strange quark within the context of an $N_f=2$ simulation 
requires the computation of mesons with valence quarks
unequal to the dynamical light sea-quarks. In addition to $\mot$ we
introduce the following generic notation:
\begin{itemize}
\item
$\moth$ - one valence quark is identical to the sea-quark.
\item
$\mtf$ - neither valence quark is identical to the sea-quark.
\end{itemize}
Furthermore, we define an effective $\kappa$ through $
{1 \over \kappa^{\rm eff}_{\rm v}} = {1 \over 2} \left( {1\over
    \kappa^1_{\rm v}} + {1 \over \kappa^2_{\rm v}}
  \right)$, where $\kappa^1_{\rm v}$ and $\kappa^2_{\rm v}$ refer to 
valence quarks in a meson. 
\par
In principle, we can fit $\mpstf$ and $\mpsoth$ to independent
linear functions in $\oks$ and $\okv$,
\bea
\mpstf &=& c_1 + c_2 \okv + c_3 \oks + \cdots \; ,  \\
\mpsoth &=&  c_1' + c_2' \okv + c_3' \oks + \cdots \; .
\label{par}
\eea
However, from the requirement that all parametrizations must converge smoothly into
each other on the symmetric line, $\ks = \kv$, the number of
independent parameters can be substantially reduced. For $\mpstf$, for
example, one finds $c_2 + c_3 = c$ and $c_1 =-{c
  \over \ksc}$. In particular, at the critical point $\mpsot = \mpsoth =\mpstf
= 0$, valence and sea quark masses must be identical:
$\kvc(\kc) = \kc$.  
This simplifies the mass equations to: 
\begin{eqnarray}
\mpstf &=& -c \okc + c_3 \oks + \okv \left( c - c_3 \right) \; ,
\\
\mpsoth &=& -c \okc + c_3' \oks + \okv \left(c -c_3' \right) \; .
\end{eqnarray}
Defining a bare valence quark mass with respect to $\ksc$ as:
\beq
\mqv = {1 \over 2}\left({1 \over \kappa_{\rm v}^{\rm eff}} - \okc
\right)\;\; ,
\label{barequarks}
\eeq
we can summarize the combined parametrization as follows\cite{Luescher}: 
\begin{eqnarray}
\left( 
\begin{array}{c} 
\mpsot  \\
\mpsoth \\
\mpstf  
\end{array}
\right) &=&  
\left( 
\begin{array}{cc} 
2 c & 0  \\
c + c_{13} & c - c_{13} \\
c_{34} & 2c - c_{34} 
\end{array}
\right)   
\left( 
\begin{array}{c} 
\mqs \\
\mqv
\end{array}
\right)   \;\; ,
\label{quarkmatrix}
\end{eqnarray}
where the variables $c_{34} = 2 c_3$ and $c_{13} = 2c_3' - c$ have been introduced.
Similarily, the vector-data can be parametrized as:
\begin{eqnarray}
\left( 
\begin{array}{c} 
\mvot  \\
\mvoth \\
\mvtf  
\end{array}
\right) &=&  m^{\rm crit} \, + \, b 
\left( 
\begin{array}{c} 
\mpsot  \\
\mpsoth \\
\mpstf  
\end{array}
\right)\;\; .
\label{quarkmatrix1}
\end{eqnarray}
%
Once the fit parameters $c$, $c_{13}$, $c_{34}$ and $\kc$ for the pseudoscalar
and additionally $m^{\rm crit}$ and $b$ for the vector particle have
been determined, we can calculate all quark masses; no additonal
fitting is needed.
\par
%
\begin{figure}[tbp]
\centerline{
\ewxynarrow{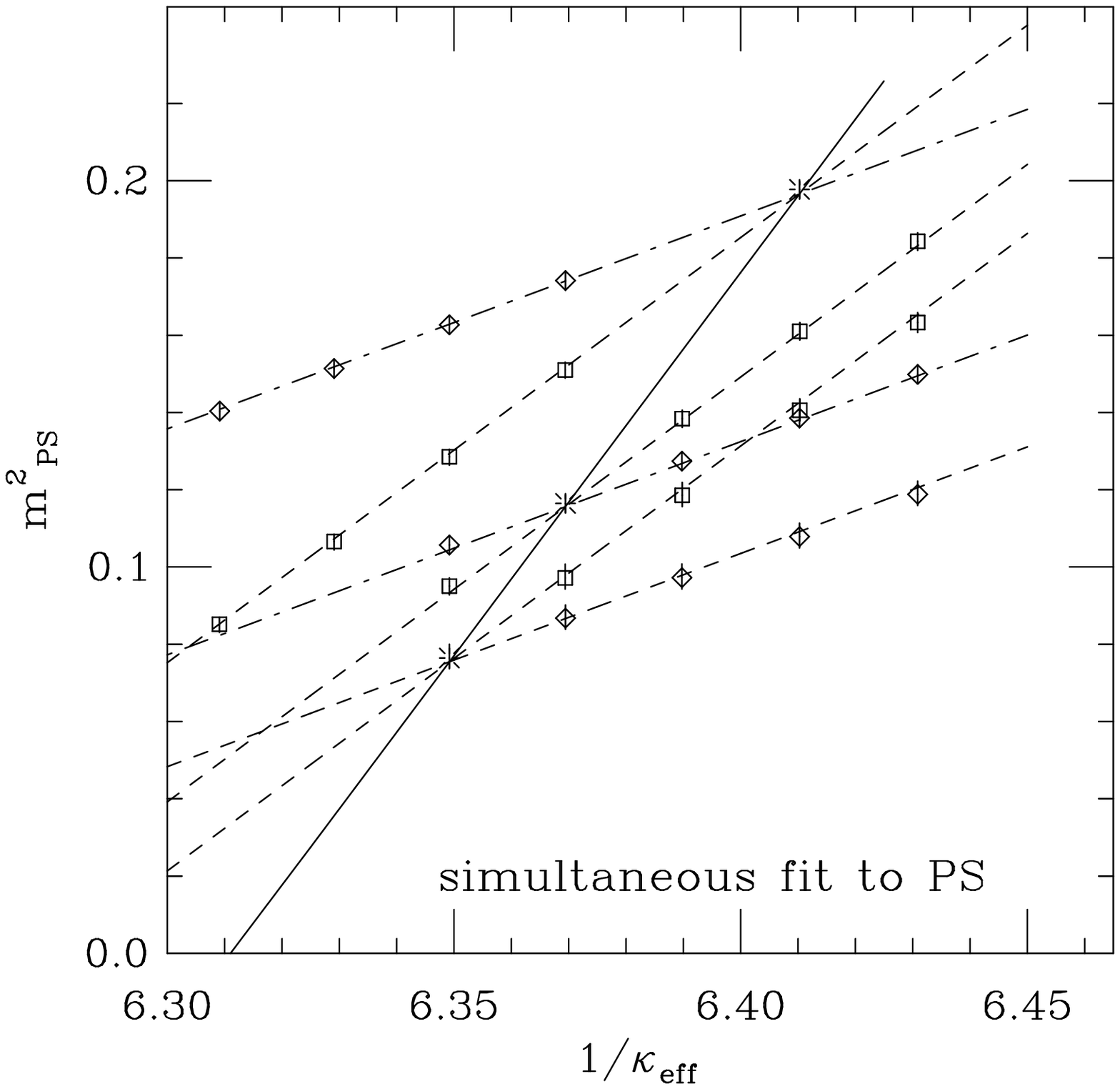}{250pt}
\ewxynarrow{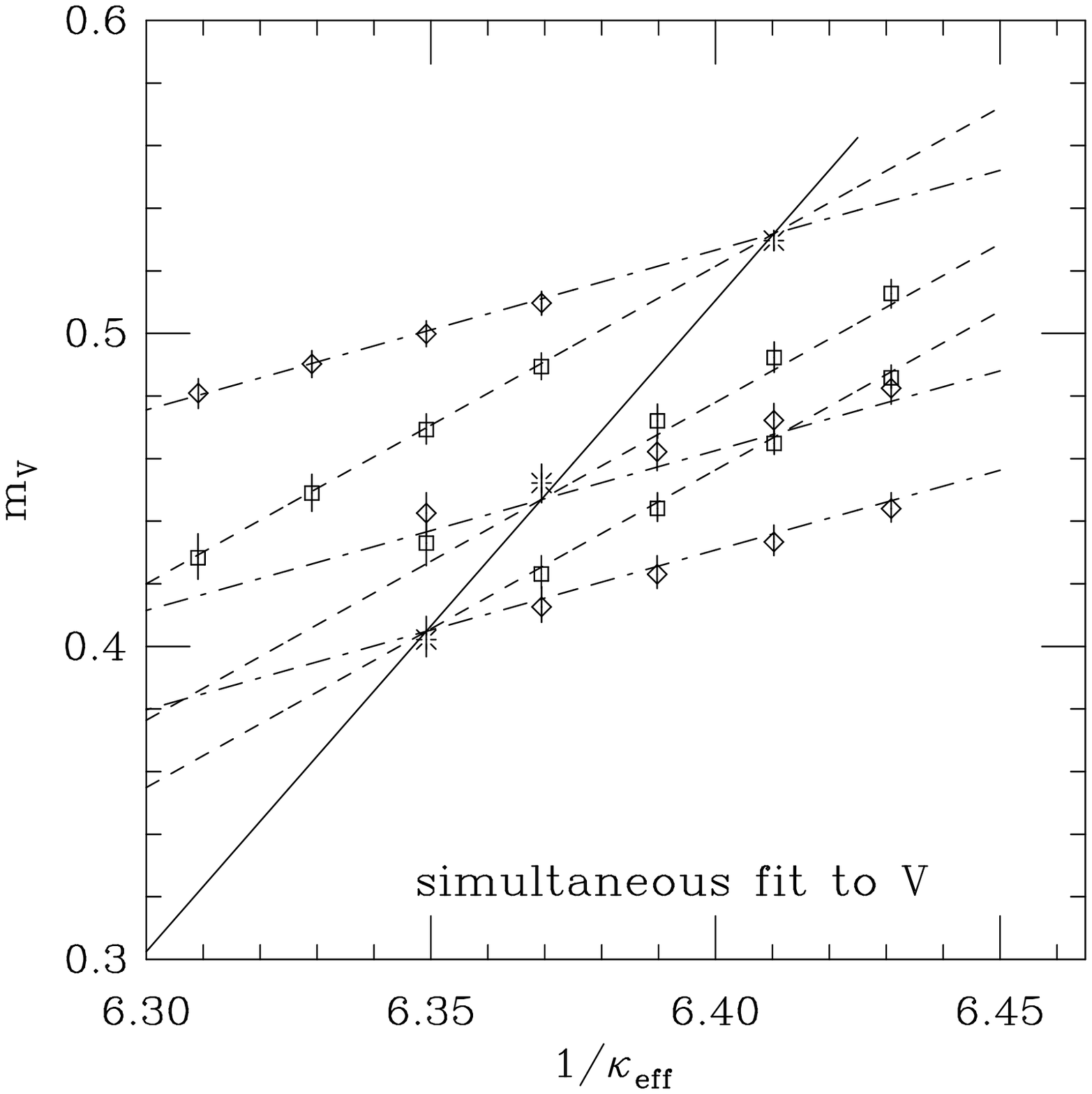}{250pt}       
}
\caption{Simultaneous fit of all pseudoscalar data to
  \eq\ref{quarkmatrix}. The vector data are fit to a semi-constrained
  form (see text). Symbols: $* = \mot\; -\!\!-$; $\Diamond = \moth \; -
  \cdot\; -$; $\Box = \mtf \; - \, -$. All plots in lattice units.
\label{hugefit}}
\end{figure}
%
Figure \ref{hugefit} (left plot) shows such a combined linear
extrapolation of all the pseudoscalar data with the ansatz of
\eq\ref{quarkmatrix}. The data are nicely fitted by this parametrization
($\chisq = 4.4/23$). A constrained fit to the vector-data with
\eq\ref{quarkmatrix1} turns out to be more difficult; we find a
$\chisq = 50/25$, fit-parameters are given in table \ref{simfits}. A
much better $\chisq$ for the vector particle can be achieved by a
semi-constrained fit in which only $\kc$, as given in \eq\ref{res1},
is held fixed, but the parameters $c$, $c_{13}$ and $c_{34}$ are
allowed to vary. This is shown in figure \ref{hugefit} on the
r.h.s.. The data for $\moth$ and $\mtf$ are best fit for those data
points of $\mot$ which match \eq\ref{pisym} well.
\par
Before we proceed to extract the strange quark mass, let us
check for the consistency of the method by using the combined
equation to determine the light quark mass and by comparing with our
previous result, eq. \ref{res2}.
%
\begin{table}[tb]
\begin{center}
\begin{tabular}{|c|c|c|c|c||c|c||c|}
\hline\hline
$\ksc$ &  $\ksl$ & $c$ & $c_{13}$ & $c_{34}$ & $b$ & $m^{\rm
  crit}$ & $\chisq$ \\ 
\hline
\multicolumn{8}{|c|}{symmetric fit to $\mpsot$ and $\mvot$}\\
\hline
$0.15846\err{5}{5}$ & $0.15841\err{5}{5}$ & $1.99\err{6}{6}$ & - & - &
$1.02\err{4}{6}$  &$0.328\err{10}{8}$ & $0.0024$, $1.1$\\ 
\hline
\multicolumn{8}{|c|}{constrained fit to \eq\ref{quarkmatrix} and \eq\ref{quarkmatrix1}}\\
\hline
$0.15845\err{5}{5}$ & $0.15841\err{5}{4}$ & $1.98\err{6}{6}$ &
$0.88\err{6}{6}$ & $1.7\err{1}{1}$ & $1.02\err{3}{5}$ &
$0.329\err{8}{6}$ & $4.4/23$, $50/25$\\ 
\hline
\multicolumn{8}{|c|}{semi-constrained fit to \eq\ref{quarkmatrix1}}\\
\hline
$\ksc$ &  $\ksl$ & $c^{\sf rho}$ & $c_{13}^{\sf rho}$ & $c_{34}^{\sf
  rho}$ & $b$ & $m^{\rm  crit}$ & $\chisq$ \\ 
\hline
$0.15845\err{5}{5}$ & $0.15841\err{5}{5}$ & $2.08\err{10}{12}$ &
$1.06\err{9}{12}$ & $2.1\err{2}{2}$ & - & $0.326\err{9}{7}$ & $4.4/23$, $6.3/23$\\
\hline\hline
\end{tabular}
\caption{Fit results from constrained (equations
  \ref{quarkmatrix} and \ref{quarkmatrix1}) as well as from symmetric
  fits to the pseudoscalar and vector and the values of $\ksc$ and $\ksl$.
\label{simfits}}
\end{center}
\end{table} 
Table \ref{simfits} shows that all three fits lead to stable values
of $m^{\rm crit}$ as well as $\kc$ and $\ksl$. This gives us
confidence in our method of choice: we employ a semi-constrained fit to
extract the strange quark mass. 
\par 
The data $\moth$ and $\mtf$ are used as follows:
\begin{itemize}
\item
Determine $\kst$ from $\moth$ by matching
\beq
{m_{\rm V, {\sf sv}}(\ksl,\kst) \over m_{\rm V, {\sf ss}}(\ksl)} = {M_{K^*} \over M_{\rho}} = 1.16 \; ,
\eeq
where $\kl$ is given by \eq\ref{res1}. Alternatively, the $K$ can be
used\footnote{In which case we match ${M_{K}^2 \over M_{\rho}^2}$.}.
\item
Determine $\kst$ from $\moth$ by matching
\beq
{m_{V_{1}, {\rm vv}}(\ksl,\kst) \over  m_{V_{2}, {\rm ss}}(\ksl)} =
  {M_{\phi} \over M_{\rho}} = 1.326 \; .
\eeq
\end{itemize}
\par
\begin{figure}[tb]
\centerline{
\ewxynarrow{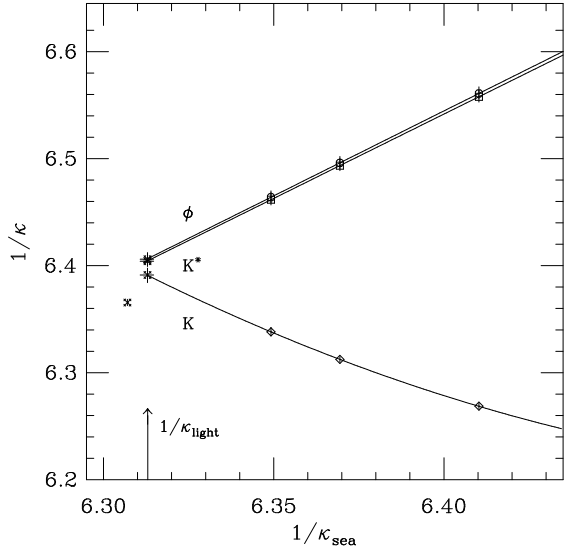}{250pt}       
}
\caption{Chiral behaviour of $\okst$, as determined from $K^*$($\Box$), $\phi$($\bigcirc$) and the
  $K$($\Diamond$). The strange quark mass is calculated with the light quark mass 
  fixed to its physical value. At the very left we plot the value for a strange
  quark in a strange sea. Lines are calculated from
  eq. \ref{quarkmatrix} and eq. \ref{quarkmatrix1}.
\label{fig5}}
\end{figure}

Table \ref{strangefits} shows the results for all three particles
employed. Good agreement is found between the $\phi$ and the $K^*$,
whereas the $K$ favours a slightly larger $\kst$. This is illustrated
in figure \ref{fig5}, where $\okst$ is plotted as a funtion of $\oks$.
\par
To check for a systematic error we have also tried constrained fits to
the vector particle. Encouragingly, we find 
only negligible change (see table \ref{strangefits}).
\par
Our final value is: 
\beq
\kst = 0.15615(20)^{\sf stat}(20)^{\sf syst}\;, 
\eeq
where the systematic error covers the spread for all three
particles. The conversion yields:
\beq
m^{\rm strange}_{\overline{MS}}(2\,{\sf GeV}) = 140(20)\, {\sf MeV} \; .
\eeq
Before we turn to a discussion of our results, we comment on a method
which has been proposed for the extraction of $\kst$ with 2 dynamical
quarks (see \cite{gupta}, for example). It circumvents the introduction of valence quarks unequal to
the dynamical quarks and determines $\kst$ by matching the ratio
${M_{\phi} \over M_{\rho}}$ using the symmetric fit only. Adopting
this procedure, one finds $\kst= 0.15709\err{8}{12}$ and $m^{\rm
  strange}_{\overline{MS}}(2\,{\sf GeV}) = 80(8)\, {\sf MeV}$; this
point is shown on the very left in figure \ref{fig5}. However, this method implies that the
$\phi$ mass is measured with strange valence quarks in a sea of
strange quarks. One would prefer the mass to be determined in a
sea of light quarks, at which, naively, one expects the effects of
dynamical sea-quarks to be more pronounced. It is interesting to see
that the effect of light sea-quarks is fairly significant: the $\phi$,
when measured in a sea of strange quarks, yields the largest of
all estimates for $\kappa^{\rm strange}$.
\begin{table}[tb]
\begin{center}
\begin{tabular}{|c|c|c|}
\hline\hline
particle &  $\kappa^{\rm strange}$ constrained & $\kappa^{\rm strange}$ semi-constrained \\
\hline
$K$ & $0.1564\err{2}{1}$ & $0.1565\err{2}{2}$ \\
\hline
$K^*$ & $0.1561\err{1}{2}$ & $0.1559\err{1}{2}$ \\
\hline
$\phi$ & $0.1561\err{1}{2}$ & $0.1559\err{1}{1}$ \\
\hline\hline
\end{tabular}
\caption{Collection of results for $\kst$.
\label{strangefits}}
\end{center}
\end{table} 
\section{Discussion\label{discussion}}
%
Our results
\begin{eqnarray}
\left. 
\begin{array}{c} 
m^{\rm light}_{\overline{MS}}(2\, {\sf GeV}) = 2.7(2)\, {\sf MeV}  \\
m^{\rm strange}_{\overline{MS}}(2\,{\sf GeV}) = 140(20)\, {\sf MeV}
\end{array}
\right\} & & \;\;\;\;\mbox{{\rm 2 dynamical light quarks}}  \; ,
\label{final1}
\end{eqnarray}
are to be compared to  the analogous quenched values at corresponding 
$\beta_{\rm quenched} = 6.0$:
\begin{eqnarray}
\left.
\begin{array}{c}
m^{\rm light}_{\overline{MS}}(2\, {\sf GeV}) = 5.5(5)\, {\sf MeV} \\
m^{\rm strange}_{\overline{MS}}(2\,{\sf GeV}) = 166(15)\, {\sf MeV}
\end{array}
\right\} & & \;\;\;\;\mbox{{\rm quenched}} \; .
\label{final2}
\end{eqnarray}
 Errors due to the finite lattice spacing and the finite volume are
 not included in (\ref{final1}) and (\ref{final2}).  
\par
Compared to the quenched result, which is in good agreement with
previous lattice calcula\-tions\cite{gupta}, we observe a
much smaller dynamical light quark mass, whereas the 
strange masses are compatible within errors. Our dynamical result for the
quark mass ratio, $m^{\rm strange}/m^{\rm light} \approx 52$, differs
significantly from chiral perturbation theory estimates\cite{leutwyler} or sum rule
results\cite{sumrules}. This may be due to the fact that the effects
of strange sea quarks can only be partially accounted for in an
$N_f=2$ simulation. 
\par
We note that an extrapolation of {\it quenched}
quark masses to $a=0$ yields much smaller values than
\eq\ref{final2}, $m^{\rm light}=3.4(5)$ and $m^{\rm
  strange}=100(23)$\cite{gupta}. Nevertheless, the ratio $m^{\rm strange}/m^{\rm
  light}$ from quenched simulations is fairly independent of the
lattice spacing $a$. This may be different in full QCD, as cutoff
effects could show up differently in sea and valence quarks. 
Dynamical results at other couplings are 
needed before one can compare results in the $a=0$ limit.
\par
To understand the drop in the mass of the light quark from $N_f=0$ to
$N_f=2$ at fixed coupling we have analysed our dynamical data in a
manner suggested by {\it quenched} lattice simulations. To this end, we have
defined a quark mass at {\it fixed} sea-quark through:
\beq
\mqv = {1 \over 2} ({1 \over \kappa_{\rm v}^{\rm eff}} - \okvc) \; .
\label{qmass1}
\eeq
Setting $\mpstf (\kvc) = 0$ at $\kvc \neq \ksc$ effectively forces an
unphysical pion to become massless. Using \eq\ref{quarkmatrix} we can
calculate the {\it quenched} critical kappa at fixed sea-quark as:
\beq
{1 \over \kappa^c_{\rm v}(\kappa_{\rm sea})} = - { c_1 + { c_3 \over
    \ks} \over c_2} \; ,
\eeq
to be compared to the {\it true} critical kappa $\okc = - {c_1 \over c_2 + c_3}$.
By construction (\eq\ref{quarkmatrix}), the values of ${1 \over
  \kappa^c_{\rm v}(\kappa_{\rm sea})}$ will lie on a straight line
hitting $\okc$ on the symmetric line.
The condition $\okvc = \okc$ has two solutions, $c_3 = 0$ and $c_3 +
c_2 = -c_1 \ks$. The latter case is identical to the true critical
kappa. The trivial condition $c_3 = 0$ corresponds to vanishing
sea-quark dependence.
\begin{center}
\begin{figure}[tb]
\epsfxsize=8cm
\epsfbox{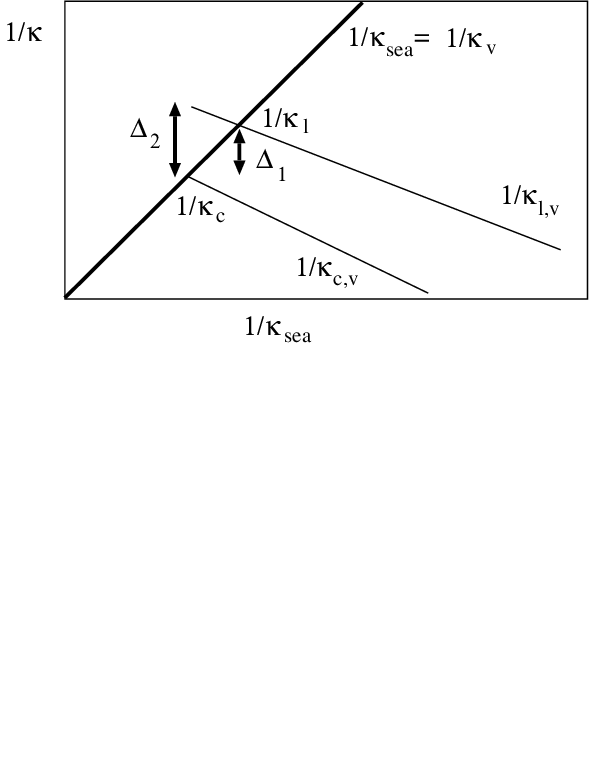}
\vspace{-5.5cm}
\caption{The values of $\okvc$ and $\okvl$ are plotted versus
  $\oks$. The situation is displayed schematically around the region of the
  symmetric line $\ks = \kv$.
\label{geom}}
\end{figure}
\end{center}
\vspace{-1cm}
%
\par
Proceeding as in quenched simulations and measuring a bare light
quark mass at each of the three sea-quark values we find light quark
masses very similar to that of the quenched simulation ($5.7(4)$,
$5.6(3)$, $5.4(3)$ {\sf MeV}) with only a slight downward trend. 
An extrapolation of these quark masses in the sea-quark
to the critical point would yield the value $\Delta_2$ in figure
\ref{geom}, while the true value is $\Delta_1$; the latter represents the
value of quark masses for a physical pion in a physical sea-quark,
whereas $\Delta_2$ is given in a sea of massless quarks. However, if
we try to repair $\Delta_2$ by extrapolating to the light 
sea-quarks we have to give up working at the physical pion mass since
then the critical kappa $\okvc$ is too low. This means that the light quark
mass, which we wish to obtain from the physical pion mass and in a sea
of physical up and down quarks, cannot be obtained by extrapolating
values obtained at fixed sea-quark to {\it either} the critical or the
light sea-quark mass. Figure \ref{geom} also illustrates that {\it quenched}
Wilson-like light quark masses away from $\okc$ should be treated with caution
since they lead to negative quark masses when measured with respect to
the physical point $\okc$, $(\okvc - \okc) < 0$.
\par
Finally, we note that a study of finite size effects is under
way\cite{tchl}. We shall include an additional
sea-quark value which will allow us to study the
effects of higher order terms in chiral perturbation theory. For the
simulation presented here, we postpone such a discussion to 
\cite{SESspec}. An analysis of the bottomonium spectrum, currently in
progress, will allow us to use a lattice spacing obtained from the
$1S-1P$ splitting, which should be less sensitive to lattice
artefacts. Much more study is needed, of course, to gain control over
these.
\vspace{0.3cm}
\\
\noindent
\underline{Acknowledgements}
\\
We wish to thank M.~L\"uscher for an interesting discussion.
%

%

\begin{thebibliography}{99}
\bibitem{leutwyler}
{H.~Leutwyler, Phys.~Lett.~ B 378, 313-318 (1996); also: hep-ph/9609467, Bern
  preprint, September 1996.} 
\bibitem{sumrules}
{For a recent analysis and early references, see J.
    Bijnens, J. Prades, and E. de Rafael, Phys.~Lett.~B348 (1994), 226.}
\bibitem{mackenzie}
{B.~J.~Gough et al., Fermilab-Pub-96-283, hep-ph/9610223; see also  the review
  of P.B. Mackenzie, Nucl.~Phys.~Proc.~Suppl.53 (1997), 23.} 
\bibitem{gupta} 
{R.~Gupta and T.~Bhattacharya, hep-lat/9605039, Phys.~Rev.~ D, to
  appear.} 
\bibitem{guptaneu}
{T.~Bhattacharya, R.~Gupta and K.~Maltman, LANL preprint
  LA-UR-96-2698, hep-ph/9703455.} 
\bibitem{Wupsmear}
{S.~G\"usken et al., Nucl.~Phys.~Proc.~Suppl.17 (1990), 361.} 
\bibitem{SESspec}
{SESAM--Collaboration, Spectrum and Decay Constants in Full QCD; in preparation.}
\bibitem{SESauto}
{SESAM--Collaboration, Performance of the Hybrid Monte Carlo for QCD
  with Wilson fermions; in
      preparation.} 
\bibitem{PRD}
{R.M. Barnett et al., Physical
      Review D54 (Paticle Data Booklet), 1 (1996). We use the
      following masses: $M_{\pi} = {1 \over 2} (M_{\pi^{\pm}} +
      M_{\pi^{0}}) = 137.3\;{\sf MeV}$; $M_{\rho} = 769\;{\sf MeV}$;
      $M_{\phi} = 1.019.4\;{\sf GeV}$; $M_{K} = {1\over 2}
      (M_{K^{\pm}} + M_{K^{0}}) = 495.68\;{\sf MeV}$; $M_{K^*} =
      892\;{\sf MeV}$.} 
\bibitem{Luescher}
{M.~L\"uscher, private communication, 24.2.97.} 
\bibitem{tchl} 
{T$\chi$L - Collaboration, L.~Conti, N.~Eicker, L.~Giusti, U.~Gl\"assner, S.~G\"usken,
      H.~Hoeber, P.~Lacock, Th.~Lippert, G.~Martinelli, F.~Rapuano,
      G.~Ritzenh\"ofer, K.~Schilling, G.~Siegert A.~Spitz, J.~Viehoff;
      Nucl.~Phys.~B 53, Proc.~Suppl. (1997) 222-224 and in preparation.}
\bibitem{Zhang}
{G.~ Martinelli and Y.~ Zhang, Phys.~ Lett.~ B {\bf 123} (1983) 433.} 
\bibitem{Lepage} 
{G.P.~Lepage and P.B.~ Mackenzie, Phys.~ Rev.~ D{\bf 48} (1993) 2250.}
\end{thebibliography}
\end{document}